\documentclass[12pt]{article}

\newbox\tempa
\newbox\tempb
\newdimen\tempc
\def\mud#1{\hfil $\displaystyle{\mathstrut #1}$\hfil}
\def\rig#1{\hfil $\displaystyle{#1}$}
\def\irulehelp#1#2#3{\setbox\tempa=\hbox{$\displaystyle{\mathstrut #2}$}%
                        \setbox\tempb=\vbox{\halign{##\cr
        \mud{#1}\cr
        \noalign{\vskip\the\lineskip}%
        \noalign{\hrule height 0pt}%
        \rig{\vbox to 0pt{\vss\hbox to 0pt{${\; #3}$\hss}\vss}}\cr
        \noalign{\hrule}%
        \noalign{\vskip\the\lineskip}%

        \mud{\copy\tempa}\cr}}%
                      \tempc=\wd\tempb
                      \advance\tempc by \wd\tempa
                      \divide\tempc by 2 }
\def\irule#1#2#3{{\irulehelp{#1}{#2}{#3}%
                     \hbox to \wd\tempa{\hss \box\tempb \hss}}}

\newcommand{\lra}{\longrightarrow}

\begin{document}

\title{From proof theory to theories theory}
\author{Gilles Dowek\thanks{\'Ecole polytechnique and INRIA.\newline
LIX, \'Ecole polytechnique,
91128 Palaiseau Cedex, France.\newline
{\tt gilles.dowek@polytechnique.edu,
  http://www.lix.polytechnique.fr/\~{}dowek}\newline
Talk given at the Lorentz center, February 12th, 2010.}}
\date{}
\maketitle
\thispagestyle{empty}

\begin{abstract}
In the last decades, several objects such as grammars, economical
agents, laws of physics... have been defined as algorithms.  In
particular, after Brouwer, Heyting, and Kolomogorov, mathematical
proofs have been defined as algorithms. In this paper, we show that
mathematical theories can be also be defined as algorithms and that
this definition has some advantages over the usual definition of
theories as sets of axioms.
\end{abstract}

\section*{The logic and the theories}

When constructing a proof, we use rules that are specific to
the objects the proof speaks about and others that are ontologically neutral.
We call the rules of the first kind {\em theoretical} rules and those
of the second {\em logical} rules.

The position of the border delimiting the logic from the theories is
not always clear and, for instance, it can be discussed if the notion
of set belongs to the realm of logic or to that of a specific
theory. It seems that the dominant point of view at the end of the
XIX$^{\mbox{th}}$ and at the beginning of the XX$^{\mbox{th}}$
century---Frege, Whitehead and Russell, ...---was that this notion of
set, or of concept, was logical. But Russell's paradox seems to have
finally ruined this point of view: as we must give up full
comprehension, and either restrict comprehension to specific
propositions, as in Zermelo set theory, or add a type system as in the
{\em Principia Mathematica}, there are too many ways to do so---with
or without the replacement axiom, with or without transfinite
types---{\em hence} the notion of set is theoretical.  We may
speculate that, even if full comprehension had not been contradictory,
this idea that the notion of set is ontologically neutral would have
been finally given up, as even with full comprehension, there still
would have been choices to be made for the set existence axioms: first
because comprehension depends on the choice of a language, then
because other set existence axioms, such as the axiom of choice may be
added.

This understanding of the theoretical nature of the notion of set has
lead, at the end of the twenties, to a separation of set theory from
predicate logic and to the constitution of predicate logic as an
autonomous object. As a corollary, it has been assumed that any theory
should be expressed as a set of axioms in predicate logic. As we shall
see, and this will be the main thesis of this paper, this 
corollary is far from being necessary.

\section*{From a historical point of view}

The constitution of predicate logic as an autonomous object,
independent of any particular theory, and the simplicity of this
formalism, compared to any particular theory such as geometry,
arithmetic, or set theory, has lead to the development of a branch
of proof theory that focuses on predicate logic. A central result in
this branch of proof theory is the cut elimination theorem: 
if a sequent $\Gamma \vdash A$ has a proof, it also has a cut-free proof, 
both in natural deduction and in sequent calculus.

Once we have a notion of proof of a sequent in predicate logic, we may
define a notion of proof of a sequent in a theory---{\em i.e.} a set
of axioms---${\cal T}$, and say that a proof of the sequent $\Gamma
\vdash A$ in the theory ${\cal T}$ is a proof in predicate logic of a
sequent $\Gamma, {\cal T}' \vdash A$, where ${\cal T}'$ is a finite
subset of ${\cal T}$.  And from results about proofs in predicate
logic, we may deduce results about proofs in arbitrary theories. For
instance, a cut elimination theorem: if a sequent $\Gamma \vdash A$
has a proof in a theory ${\cal T}$, it has a cut free proof in this
same theory.

Yet, although such results do have some content, they often are
disappointing: they do not say as much as we could expect. For
instance, the cut free proofs of a sequent $\vdash A$ in predicate
logic always end with an introduction rule, but this in not the case for the
cut free proofs in a theory ${\cal T}$. Thus, while a corollary of the
cut elimination theorem for predicate logic is that there is no proof
of the sequent $\vdash \bot$ in predicate logic, this corollary does not
extend to arbitrary theories.  And, indeed, there are contradictory
theories.

Thus, this definition of a cut free proof of a sequent $\Gamma \vdash
A$ in a theory ${\cal T}$ as a cut free proof of a sequent
$\Gamma, {\cal T}' \vdash A$ for some finite subset ${\cal T}'$ of
${\cal T}$ is not satisfactory, as the existence of such cut free
proofs does not imply the consistency of the theory. In the
same way, it does not imply the disjunction or the witness property
for constructive proofs.

Thus, this approach to proof theory focused on predicate logic
has been unable to handle theories in a satisfactory way. This may
be a sign that the definition of the notion of theory as a set of
axioms is too general: not much can be said about a too large class of
objects and a more restrictive notion of theory might allow to say
more about theories.

From a historical point of view, proof theory has side stepped this
problem in at least three ways. First, some results, such as cut
elimination theorems, have been proven for specific theories, for
instance arithmetic. Virtually, any set of axioms yields a new branch of
proof theory.  Then, the separation between the logic and
the theories has been criticized and extended logics have been
considered. Typical examples are second-order logic, higher-order
logic, Intuitionistic type theory \cite{MartinLof}, the Calculus of 
constructions \cite{CoquandHuet}, ...
Finally, the idea that the proofs studied by proof theory are proof of
mathematical theorems, and thus require a theory, has be given up and
proofs have been studied for for their own sake. A typical example is
linear logic \cite{GirardLinear}.

The thesis we shall develop in this paper is that there is another
possible way to go for proof theory: modify the notion of theory so
that it can be properly handled, without focusing on a single theory,
such as arithmetic, and without introducing logics beyond
predicate logic, such as second-order logic.  This way leads to a
branch of proof theory that may be called {\em the theory of
theories}.

\section*{Some technical problems with axioms}

We have seen that the cut elimination theorem for predicate logic
implies that there is no proof of the sequent $\vdash \bot$ in
predicate logic, but that this result does not extend to an arbitrary
theory. Let us now consider a more evolved example.

A consequence of the cut elimination theorem is that a bottom up proof
search in sequent calculus is complete, even if it is restricted to
cut free proofs. In the search for a proof of the sequent $\vdash
\bot$, no rule of the cut free sequent calculus applies---except
structural rules in some versions of the sequent calculus---and the
search fails in finite time. In contrast, the search for a cut free
proof of the sequent $\vdash \bot$ in the theory $\forall x~(P(x)
\Leftrightarrow P(f(x)))$, {\em i.e.} for a cut free proof of the sequent
$\forall x~(P(x) \Leftrightarrow P(f(x))) \vdash \bot$ in predicate
logic, generates an infinite search space. Thus, although the cut
elimination theorem allows to reduce the size of the search space, by
restricting the search to cut free proofs, it does not allow to reduce
it enough so that the search for a proof of a contradiction in the
theory $\forall x~(P(x) \Leftrightarrow P(f(x)))$ fails in finite
time.  This proof search method ``does not know'' \cite{CADE}
that this theory is
consistent and indeed the cut elimination theorem for predicate logic
does not imply, {\em per se}, the consistency of any non-empty set of 
axioms.

In order to design a complete proof search method for this theory, that 
fails in finite time when searching for a proof of $\vdash \bot$, 
we may attempt to prove a cut elimination theorem, specialized for this
theory, that implies its consistency, in the same way as the cut elimination 
theorem for arithmetic implies the consistency of arithmetic.

However, when attempting to follow this path, the first problem we
face is not to prove this cut elimination theorem, it is to state it:
to define a notion of cut specialized for this theory.  Indeed,
although there is an {\em ad hoc} notion of cut associated to the
axioms of arithmetic---a sequence where a proposition is proven by
induction and then used in a particular case---there is no known way
to associate a notion of cut to an arbitrary set of axioms.

Among the tools used in proof theory, only one generalizes smoothly
from predicate logic to arbitrary sets of axioms: the notion of model.
Indeed, not only the definition of the notion of model extends from
predicate logic to arbitrary sets of axioms, but the completeness
theorem also does.  In this respect, there is a clear contrast between
the notion of model and that of cut. And the definition of the notion
of theory as a set of axioms appears to be tailored for the notion of
model and not for that of cut.

\section*{Two partial solutions}

There are several cases where this problem of the definition of the
notion of cut associated to a theory has been addressed and where partial 
solutions have been proposed. When enumerating these cases, we also have 
to include cases where a proof search method has been proven complete. 
Indeed, as shown by Olivier Hermant \cite{Hermant}, the completeness of a proof 
search method is often not only a consequence of a cut elimination theorem,
but it is also equivalent to such a theorem.

Let us start with a slightly artificial example: that of definitions. If
we want to use a new proposition symbol $P$ as an abbreviation for a
proposition $A \wedge B$, we can either keep the same language and 
consider each occurrence of the symbol $P$ as a notational variant for
$A \wedge B$, or extend the language with the symbol $P$ and add the axiom 
$P \Leftrightarrow (A \wedge B)$.
Obviously, the same propositions can be proven in both cases and from
the cut elimination theorem for predicate logic, we should be able to
derive the consistency of the empty theory, and hence that
of the theory $P \Leftrightarrow (A \wedge B)$. In the same way,
from the cut elimination theorem for predicate logic, we should be able to 
derive the completeness of a proof search method that
first uses the axiom $P \Leftrightarrow (A \wedge B)$ to replace all
the occurrences of $P$ by $A \wedge B$ in the proposition to be proven 
and then searches for a cut free proof of the obtained proposition.

A direct cut elimination proof for this theory can be given following 
an idea of Dag Prawitz \cite{Prawitz}. In constructive natural deduction, 
we can 
replace the axiom $P \Leftrightarrow (A \wedge B)$ by two {\em theoretical} 
deduction rules 
$$\irule{\Gamma \vdash A \wedge B}{\Gamma \vdash P}{\mbox{fold}}$$
and 
$$\irule{\Gamma \vdash P}{\Gamma \vdash A \wedge B}{\mbox{unfold}}$$
and extend the notion of cut in such a way that the sequence of a {\em
fold} rule and an {\em unfold} rule is a cut.  Cut elimination can
be proven for this theory and like in predicate logic, a cut free
proof always end with a introduction rule---considering the {\em
  fold} rule as an introduction rule and the {\em unfold} rule as an
elimination rule.  Thus, all the corollaries of cut elimination that
are based on the fact that the last rule of a cut free proof is an
introduction rule, such as consistency, the disjunction property, the
witness property, ... generalize.

Of course, there is nothing specific to the proposition $A \wedge B$
in this example and we can do the same thing with any theory of the
form $P_1 \Leftrightarrow A_1, ..., P_n \Leftrightarrow A_n$ where
$P_1, ..., P_n$ are proposition symbols that do not occur in $A_1$,
..., $A_n$. This class of theories is quite small and it is fair to
say that all these theories are trivial. Yet, we may notice that a
notion of cut has been defined for a class of theories that contains more
than just one theory.

Besides this quite artificial example of definitions, this idea of
using theoretical deduction rules 
has been investigated by Dag Prawitz \cite{Prawitz},
Marcel Crabb\'e \cite{Crabbe74,Crabbe91}, Lars 
Halln\"as \cite{Hallnas}, Jan Ekman 
\cite{Ekman}, Sara Negri and Jan von Plato \cite{NegriPlato}, ...

More precisely, the {\em fold} and {\em unfold} rules can be used each time 
we have a theory of the form 
$\overline{\forall} (P_1 \Leftrightarrow A_1), ..., \overline{\forall} 
(P_n \Leftrightarrow A_n)$ where $P_1, ..., P_n$ are arbitrary atomic 
propositions.
In particular, this can be applied successfully to set 
theory where, for instance, the axiom 
$\forall x \forall y~(x \in {\cal P}(y) \Leftrightarrow \forall z~(z \in x \Rightarrow z \in y))$ 
can be replaced by two rules 
$$\irule{\Gamma \vdash \forall z~(z \in x \Rightarrow z \in y)}
        {\Gamma \vdash x \in {\cal P}(y)}
        {\mbox{fold}}$$
and 
$$\irule{\Gamma \vdash x \in {\cal P}(y)}
        {\Gamma \vdash \forall z~(z \in x \Rightarrow z \in y)}
        {\mbox{unfold}}$$

A remark due to Marcel Crabb\'e \cite{Crabbe74} is that
there are theories for which cut elimination does not hold. For
instance with the rules
$$\irule{\Gamma \vdash P \Rightarrow Q}{\Gamma \vdash P}{\mbox{fold}}$$
and 
$$\irule{\Gamma \vdash P}{\Gamma \vdash P \Rightarrow Q}{\mbox{unfold}}$$
we have a proof of $Q$ but no cut free proof of $Q$.  Yet, whether or
not cut elimination holds, the notion of cut can be defined and the
cut elimination problem can be stated for all these theories.

A second example comes from an attempt to handle equality in automated 
theorem proving. It has been remarked for long that considering 
equality axioms such as 
$$\forall x \forall y \forall z~(x + (y + z) = (x + y) + z)$$
together with the 
axioms of equality generates a very large search space, even 
when this search is
restricted to cut free proofs.
For instance, attempting to prove the proposition 
$$P((a + b) + ((c + d) + e)) \Rightarrow P(((a + (b + c)) + d) + e)$$
we may move brackets left and right  in many ways, using the associativity 
axiom and the axioms of equality, before reaching a proposition that has the 
form $A \Rightarrow A$. An idea, that goes back to Max Newman \cite{Newman},
but that has been fully developed by Donald Knuth and Peter Bendix \cite{KB}, 
is to 
replace 
the associativity axiom by the rewrite rule 
$$x + (y + z) \lra (x + y) + z$$
that allows to transform the proposition $P((a + b) + ((c + d) + e))$ into
$P((((a + b) + c) + d) + e)$ but not vice-versa. If the rewrite rules
form a terminating and confluent system, then two terms are provably
equal if and only if they have the same normal form. 
Normalizing the proposition above 
yields the provably equivalent proposition
$$P((((a + b) + c) + d) + e) \Rightarrow P((((a + b) + c) + d) + e)$$
that is provable without the associativity axiom.

Yet, even if the rewrite system is confluent and terminating, normalizing 
propositions at all times is not sufficient to get rid of
the associativity axiom. 
Indeed, the proposition
$$\exists x~(P(a + x) \Rightarrow P((a + b) + c))$$
is provable using the associativity axiom, but its normal form,
{\em i.e.} the proposition itself, is not provable without the
associativity axiom. In particular, the propositions $P(a + x)$ and
$P((a + b) + c)$ are not unifiable, {\em i.e.} there is no substitution of
the variable $x$ that makes these two propositions identical.  The
explanation is that, in the proof of this proposition, we must first
substitute the term $b + c$ for the variable $x$ and {\em then}
rewrite the proposition $P(a + (b + c)) \Rightarrow P((a + b) + c)$ to
$P((a + b) + c) \Rightarrow P((a + b) + c)$ before we reach a
proposition of the form $A \Rightarrow A$. To design
a complete proof search method, the unification algorithm must be
replaced by an algorithm that searches for a substitution that does
not make the two propositions equal, but let them have the same normal
form.  Such an algorithm is called {\em an equational unification
algorithm modulo associativity}. And Gordon Plotkin \cite{Plotkin}
has proven that when unification is replaced by equational unification
modulo associativity, the associativity axiom can be dropped without
jeopardizing the completeness of the proof search method.

This proof search method is much more efficient than the method that
searches for a cut free proof in predicate logic using the
associativity axiom and the axioms of equality. In particular, the
search for a cut free proof of the sequent $\vdash \bot$ with
equational unification fails in finite time, unlike the search of a
cut free proof of the sequent $\vdash \bot$, using the associativity
axiom and the axioms of equality.

When formulated, not in a proof search setting, but in a purely proof
theoretical one, Plotkin's idea boils down to that that two
propositions such as $P(a + (b + c))$ and $P((a + b) + c)$ that have
the same normal form, should be identified. We have called {\em
  Deduction modulo} \cite{DHK,DW} this idea to reason modulo 
a congruence
on propositions---{\em i.e.} an equivalence relation that is compatible
with all the symbols of the language.  Then, the need to replace
unification by equational unification is simply a consequence of the
fact that when unifying two propositions $P$ and $Q$, we must find a
substitution $\sigma$ such that $\sigma P$ and $\sigma Q$ are
identical, modulo the congruence.  The possibility to drop the
associativity axiom is just a consequence of the fact that this axiom
is equivalent, modulo associativity, to the proposition $\forall x
\forall y \forall z~((x + y) + z = (x + y) + z)$ that is a consequence
of the axiom $\forall x~(x = x)$. Finally, the completeness of proof
search modulo associativity is a consequence of the fact that, when 
terms are identified modulo associativity in predicate logic, cut
elimination is preserved because cut elimination ignores the inner 
structure of atomic
propositions. Thus, a cut in Deduction modulo is just a sequence formed with
an introduction rules and an elimination rule, like in predicate logic, 
the only difference being that these rules are applied modulo 
a congruence.

This remark generalizes to all equational theories, {\em i.e.} to all 
the theories whose axioms have the form $\overline{\forall}~(t = u)$
and we get this way a cut elimination theorem for a class of 
theories that contains more than one theory and from this 
cut elimination theorem, we can rationally reconstruct the
proof search methods for equational theories designed by Gordon Plotkin.

\section*{Deduction modulo}

Deduction modulo is thus a tool that permits to associate a notion of
cut to any equational theory.

But this notion of cut also subsumes that introduced with the {\em
fold} and {\em unfold} rules. Indeed, if we start with a theory formed
with the axioms $\overline{\forall}~(P_1 \Leftrightarrow A_1), ...,
\overline{\forall}~(P_n \Leftrightarrow A_n)$, instead of replacing
these axioms by theoretical deduction rules, we can replace them by
rewrite rules $P_1 \lra A_1, ..., P_n \lra A_n$ that rewrite
propositions directly---and not only terms that occur in
propositions---and consider the congruence defined with these rules.
We obtain this way a notion of cut that is equivalent to the notion of
cut introduced with the {\em fold} and {\em unfold} rules: not all
theories have the cut elimination property, but a theory has the cut
elimination property in one case if and only if it has this property
in the other.  Although it is not difficult to prove, this result is
not completely trivial because Deduction modulo allows some deep
inferences---a rewrite rules can be used at any place in a
proposition, while the {\em fold} and {\em unfold} rules allow shallow
inferences only, {\em i.e.}  rewriting at the root, and this is essential
for the cut elimination to work with the {\em fold} and {\em unfold}
rules.

By an abuse of language, when Deduction modulo a congruence has the
cut elimination property, we sometimes say that the congruence itself 
has the cut elimination property.

Thus, Deduction modulo is framework where a theory is not defined as a set 
of axioms but as a set of rewrite rules and this framework unifies the 
notions of cut defined by Dag Prawitz and by Gordon Plotkin. 

\section*{From axioms to algorithms}

An important question about Deduction modulo is how strong the
congruence can be. For instance, can we take a congruence
such that $A$ is congruent to $\top$ if $A$ is a theorem of
arithmetic, in which case each proof of each theorem is a proof of all
theorems? This seems to be a bad idea, for at least two reasons. First, 
the decidability of proof checking requires that of the congruence. 
Then, for the cut free proofs
to end with an introduction rule, the congruence must be 
{\em non-confusing}, 
{\em i.e.} when two non-atomic propositions are congruent, they 
must have the same head symbol and their sub-parts must be congruent.
For instance, if both propositions 
are conjunctions, $A \wedge B$ and $A' \wedge B'$, 
$A$ must be congruent to $A'$ and $B$ to $B'$. 

Both properties are fulfilled when the congruence is defined by a
confluent and terminating rewrite system where the left hand side of each rule 
is either a term or an atomic proposition, such as $P(0) \lra \forall
x~P(x)$.  But, we sometimes consider cases where the congruence is defined
by a rewrite system that is not terminating, for instance if it
contains the rule $x + y \lra y + x$ or the rule $P \lra P \Rightarrow
Q$. Nevertheless, the congruence defined this way may still be
decidable and non-confusing.

Cut elimination is a third property that may be required in the definition of 
the notion of theory.

As the rewrite rules define a decidable congruence, we may say that
they define an algorithm. And indeed, we know that the rewrite rules
that replace the axioms of the addition
$$0 + y \lra y$$
$$S(x) + y \lra S(x+y)$$
define an algorithm for addition.

Thus, Deduction modulo is also a way to separate the deduction part
from the computation part in proofs. This idea, formulated here in
the general setting of predicate logic had already been investigated
in some specific systems such as Intuitionistic type theory \cite{MartinLof} 
and the Calculus of
constructions \cite{CoquandHuet}, where a decidable {\em definitional
  equality} is distinguished from the usual propositional equality.
Also, the idea of transforming theorems into rewrite rules has been
used by Robert S. Boyer and J. Strother Moore in the theorem prover ACL
\cite{BoyerMoore}.

But the novelty of Deduction modulo is that rewrite rules are used as a
substitute for axioms and that provability with rewrite rules is
proven to be equivalent to provability with axioms. Thus, the rewrite
rules are used to express the theory, while the theory is not
distinguished from the logic in Intuitionistic type theory, in the
Calculus of constructions or in the Boyer-Moore logic.

Thus, it seems that Deduction modulo gives an original answer to the
question ``What is a theory ?'' and this answer is ``An algorithm''.

This answer is part of a general trend, in the last decades, to answer
``An algorithm'' to various questions, such as ``What is a grammar
?'', ``What is an economical agent ?'', ``What is a law of physics
?'', ``What is a proof ?'', ... and this answer may be given to more
questions in a near future, e.g. to the question ``What is a cell ?''

It must be noticed, however, that the idea that a proof is an
algorithm, the Brouwer-Heyting-Kolomogorov interpretation of
proofs, may have somehow been an obstacle to our understanding that a
theory also is an algorithm. The success of the algorithmic
interpretation of proofs seems to have implicitly promoted the idea that
this algorithmic interpretation of proofs was {\em the} link between
the notion of algorithm and that of proof, making it more difficult to
remark this other link.

But in the same way computer scientists know that a program expresses
an algorithm and that the algorithm used to check that is program is
correctly typed is another algorithm, it seems that proofs are algorithms
and that their correctness criterion is parametrized by another algorithm: the
theory.

\section*{Extended logics}

This new answer to the question ``What is a theory?'' gives the
possibility to define a general notion of cut for all the theories defined
in this way and to prove general cut elimination results that apply to
large classes of theories.

It gives the possibility to avoid two of the strategies mentioned
above to side step the inability of proof theory of predicate logic
to handle theories in a satisfactory way: 
introduce extended logics, such as
second-order or higher-order logic, and 
focus on particular theories, such as arithmetic. 

Second-order and higher-order logics have been studied thoughtfully in
the second half of the XX$^{\mbox{th}}$ century. Higher-order logic has a
special notion of model and a special completeness theorem \cite{Henkin}, 
a special cut elimination theorem \cite{Girard}, 
special proof search algorithms \cite{Andrews,Huet73}, 
even a special notion of substitution and a special Skolem theorem 
\cite{Miller}.

The history of the notion of higher-order substitution in itself could
have lead to Deduction modulo. Before the modern view that
substituting the term $\lambda x~(Q(f(x)))$ for the variable $P$ in
the proposition $(P~0)$ yields the proposition $(\lambda
x~(Q(f(x)))~0)$,
that is provably equivalent to $Q(f(0))$, using some conversion axioms, this
substitution yielded directly the proposition $Q(f(0))$. Thus, normalization
was included in substitution operation. 

Then, the substitution operation was simplified to an operation that is
almost the substitution of predicate logic, and conversion axioms were
introduced in the formulation of higher-order logic given by 
Alonzo Church \cite{Church}. Shortly after, the idea of incorporating
these axioms in the unification algorithm was proposed by Peter
Andrews \cite{Andrews} and this yielded higher-order unification
\cite{Huet75}.
Although Andrews' idea seems to be independent from Plotkin's, 
higher-order unification is equational 
unification modulo beta-conversion. 

Finally, in Deduction modulo, the conversion axioms are included in
the congruence and unification is performed modulo this congruence.

The story is just slightly more complex because besides
separating substitution from conversion,
Church's formulation of higher-order logic introduces a symbol
$\lambda$ that binds a variable, thus the substitution operation in
this logic must handle binders. In the same way, 
besides being equational, higher-order unification is also
nominal \cite{UPG}, {\em i.e.} it involves terms containing binders.

Yet, lambda bound variables can be eliminated from higher-order logic,
for instance using combinators or de Bruijn indices and explicit
substitutions and the last idiosyncrasy of higher-order logic, the
absence of separation between terms and propositions can be avoided by
introducing a ``unary copula'' $\varepsilon$ transforming a term of
type $o$ into a propositions. This way, higher-order logic can be
defined as a theory in Deduction modulo \cite{DHK2}, and in this case
it is better to call it {\em Simple type theory}, to stress its
theoretical nature. We avoid this way the need of a special notion of
model, a special completeness theorem, a special cut elimination
theorem \cite{DW}, special proof search algorithms \cite{DHK,polar}, a
special notion of substitution and a special Skolem theorem
\cite{AndrewsFest}. Some formulations of second-order logic, such as
the {\em Functional second-order arithmetic} of Jean-Louis Krivine and
Michel Parigot \cite{Leivant,KrivineParigot}, even take more advantage
of the notion of congruence when expressed as theories in Deduction
modulo \cite{AF2}.

All the theories expressed in Deduction modulo verify an extended
sub-formula property, and so does higher-order logic. How is this
compatible with the well-known failure of the sub-formula property for
higher-order logic?  First, we have to notice that, strictly speaking,
the sub-formula property holds for propositional logic only. The
notion of sub-formula has to be adjusted for the sub-formula property
to hold for predicate logic: the set of sub-formulae of a proposition
has to be defined as the smallest containing this proposition, that is
closed by the sub-tree relation and also by substitution. This way, the
proposition $\forall x~P(x)$ has an infinite number of sub-formulae as
all the instances of the proposition $P(x)$ are sub-formulae of the
proposition $\forall x~P(x)$.  In Deduction modulo, besides being
closed by the sub-tree relation and substitution, this set must be
closed by the congruence. Thus, if we have a rule $P \lra Q
\Rightarrow Q$, the sub-formulae of the proposition $P$ are $P$ and
$Q$. Using this definition, the set of sub-formulae of some
propositions in higher-order logic contains all the propositions.
This is what is usually called the failure of the sub-formula property
for higher-order logic.

This ability of predicate logic to handle theories changes the
organization of proof theory as a field of knowledge by giving back
predicate logic a central place and reformulating in a wider setting
the results specially developed for second-order and higher-order
logic.

\section*{Particular theories}

In the same way, the results developed for a particular theory, such
as arithmetic are just consequences of general results, once these
theories have been expressed as algorithms. And several theories have
been expressed in this way, in particular arithmetic \cite{Peano,Allali} 
and some versions of set theory \cite{Zermodulo}, although some other 
theories, such as geometry, have not been investigated yet. 

The case of arithmetic is interesting as it shows how expressing a
theory as an algorithm sheds a new light on this theory. There are two
ways to express arithmetic in predicate logic, either, as Peano did,
using a predicate $N$ characterizing natural numbers, or not.  That a
natural number is either $0$ or a successor is expressed in the first
case by the proposition
$$\forall x~(N(x) \Rightarrow (x = 0 \vee \exists y~(N(y) \wedge x = S(y))))$$
and in the second by the proposition 
$$\forall x~(x = 0 \vee \exists y~(x = S(y)))$$
This simpler formulation is usually preferred. Yet, the
formulation of arithmetic in Deduction modulo does use the Peano
predicate $N$. This formulation was originally considered as a first 
step, before we understood how to eliminate this predicate. Yet, it appears
that this predicate cannot be eliminated. Indeed, the formulation of 
arithmetic without this predicate has a strange property: in the 
constructive case, it verifies the closed disjunction property 
but not the open one. That is, if a closed proposition of the form 
$A \vee B$ is provable, then either $A$ or $B$ is provable, but 
some open propositions of the form 
$A \vee B$ are provable, while neither $A$ nor $B$ is, e.g. 
$x = 0 \vee \exists y~(x = S(y))$. And any theory expressed in 
Deduction modulo that enjoys the cut elimination property verifies the 
full disjunction property.

This means that not all sets of axioms can be transformed into a
non-confusing congruence
that has the cut elimination property.  And this is a strength, not a weakness, of Deduction
modulo. Deduction modulo rules out some theories, e.g. contradictory
theories, theories that do not verify the disjunction property in the
constructive case, ...  and the counterpart of this is that strong
results can be proven about the theory that can be expressed in
Deduction modulo, such as consistency or the disjunction property in
the 
constructive
case. The lack of strong results in the theory of theories when
theories were defined as sets of axioms is a consequence of the fact
that this notion of theory was too general, and not much can be said
about a too large class of objects.

This remark raises an difficult but interesting question. What are the
properties that a set of axioms must fulfill to be transformed into a
non-confusing congruence having the cut elimination property?  A
recent unpublished result of Guillaume Burel \cite{Burel} suggests
that, in the classical case, up to skolemization, consistency is a 
sufficient condition. In the
constructive case, the problem is open.  As all theories expressed by a
non-confusing congruence that has the cut elimination property are
consistent, and enjoy the disjunction and the witness property, these
conditions are necessary for a theory to be expressed as a non
confusing congruence that has the cut elimination property, and for
instance, the axiom $P \vee Q$ and the axiom $\exists x~P(x)$
cannot. Whether these conditions are sufficient is open.

\section*{First chapter of a theory of theories}

Deduction modulo has already lead to general cut elimination results
\cite{DW} that subsume cut elimination results for arithmetic and for
higher-order logic. It also has lead to proof search algorithms that
subsume both equational resolution and higher-order resolution
\cite{DHK,polar} and to the development of an extended notion of
model, where two congruent propositions are interpreted by the same
truth value, but not necessarily two provably equivalent propositions
\cite{TVA}.  It has also lead to a formulation of a lambda-calculus
with dependent types called $\lambda \Pi$-modulo
\cite{CousineauDowek}, that permits to express proofs in Deduction
modulo as algorithms and that subsumes the Calculus of constructions
and more generally all functional Pure type systems \cite{Barendregt}.
The important point with this calculus is that it completely
de-correlates the issue of the functional interpretation of proofs
from the problem of the choice of a theory.  The functional
interpretation of proofs is handled by the dependent types and the
theory by the congruence. There is no special link between some typing
disciplines---such as polymorphism---and some theories---such as the
second-order comprehension scheme.

But, the most interesting about Deduction modulo is that it permits to
formulate problems that could not be formulated when theories were
defined as sets of axioms. In particular, we may seek for a
characterization of the theories that have the cut elimination
property or of those that have the proof normalization property.
Olivier Hermant has given examples showing that cut elimination does not
imply normalization \cite{Hermantthese}. Then, Denis Cousineau \cite{Cousineau} 
has given the first condition
both sufficient {\em and necessary} for normalization.  The fact that
this condition is model theoretic suggests that the notion of
normalization is also a model-theoretic notion. 

This problem and the problem of the characterization of the theories 
that can be expressed in Deduction modulo are two among the main open 
problems in this area. Defining a nominal Deduction modulo, {\em i.e.} 
including 
function symbols that may bind variables is also an important one.

If we look at this theory of theories from outside, we see that
some aspects that were minor in the proof theory of predicate logic
become majors aspects: the link between the notion of cut and that of
model, the link between proof theory and automated theorem proving, 
and also many-sorted predicate logic, as unlike axioms, rewrite rules are 
difficult to relativize, ...

On the other hand, some topics that have been central in proof theory in 
the last decades become less central: the case of higher-order logic and
the functional interpretation of proofs. These notions are not given
up, but, like the notion of triangle in geometry, they just lose their 
central place by being included in a larger picture.

\end{document}